\documentclass[sigconf,nonacm]{acmart}
\AtBeginDocument{%
  }

\begin{document}

\title{Simulation-based Optimization for Augmented Reading}

\author{Yunpeng Bai}
\orcid{0009-0008-7578-0079}
\affiliation{%
  \institution{National University of Singapore}
  \country{Singapore}}

\author{Shengdong Zhao}
\orcid{0000-0001-7971-3107}
\affiliation{%
  \institution{City University of Hong Kong}
  \city{Hong Kong}
  \country{China}}

\author{Antti Oulasvirta}
\orcid{0000-0002-2498-7837}
\affiliation{%
  \institution{Aalto University}
  \city{Helsinki}
  \country{Finland}}

\renewcommand{\shortauthors}{Bai et al.}

\begin{abstract}
Augmented reading systems aim to adapt text presentation to improve comprehension and task performance, yet existing approaches rely heavily on heuristics, opaque data-driven models, or repeated human involvement in the design loop. We propose framing augmented reading as a simulation-based optimization problem grounded in resource-rational models of human reading. These models instantiate a simulated reader that allocates limited cognitive resources, such as attention, memory, and time under task demands, enabling systematic evaluation of text user interfaces. We introduce two complementary optimization pipelines: an offline approach that explores design alternatives using simulated readers, and an online approach that personalizes reading interfaces in real time using ongoing interaction data. Together, this perspective enables adaptive, explainable, and scalable augmented reading design without relying solely on human testing.
\end{abstract}



\keywords{Reading, Modeling, Optimization, User Interface}


\maketitle

\section{Introduction}

Augmented reading aims to adapt text presentation to readers’ contexts and capabilities, for example under multitasking conditions~\cite{bai2024heads,lingler2024supporting}, time pressure~\cite{bai2025resource}, or varying cognitive abilities. Recent systems have explored adaptive summaries, dynamic layouts~\cite{lindlbauer2019context}, and AI-assisted transformations of text. However, most existing approaches~\cite{reichle2003ez,salvucci2001integrated,legge1997mr} rely either on heuristic, rule-based adaptations that do not scale across contexts, or on data-driven models~\cite{bolliger2023scandl,Bolliger2025ScanDL2,deng2023eyettention,jiang2024eyeformer,jiang2024graph4gui} that require large amounts of training data and offer limited interpretability. As a result, current augmented reading systems struggle to remain simultaneously flexible, cognitively grounded, and human-centered due to the limitation of existing computational reading models.

We propose that computationally resource-rational models of reading~\cite{bai2024heads,bai2025resource} can serve as principled optimization objectives for augmented reading systems. These models conceptualize reading as adaptive decision-making under cognitive and environmental constraints, enabling the simulation of human-like reading behavior across wide situations and reader profiles~\cite{lieder2020resource,oulasvirta2022computational}. Rather than directly predicting interface adaptations from data, which often lack explicit representations of human cognition~\cite{jiang2024eyeformer,jiang2022computational,jiang2023future,jiang2023ueyes,jiang2024graph4gui,bolliger2023scandl,Bolliger2025ScanDL2,deng2023eyettention}, simulated readers can be used to evaluate and compare candidate text presentations from a human-centered perspective, accounting for comprehension, cognitive workload, and time limitations~\cite{bai2025resource,bai2024heads}. This reframes augmented reading as a simulation-based optimization problem, where simulated readers represent human readers to act as participants and involve in scalable, adaptive, and explainable text interface design. In this paper, we propose how such reading models can be integrated into offline and online optimization pipelines, and demonstrate their use through representative augmented reading scenarios.

\section{Background: Resource-rational Reading Models}

Resource-rational reading models conceptualize human reading as adaptive decision-making under cognitive and environmental constraints~\cite{bai2024heads,bai2025resource}. Rather than treating reading as a fixed sequence of perceptual and linguistic stages, these models view reading behavior as emerging from the optimization of task performance given limited resources such as attention, memory, and time. This perspective follows the broader framework of computational rationality~\cite{oulasvirta2022computational,lieder2020resource,howes2009rational,howes2016rationality,lewis2014computational}, which explains human behavior as approximately optimal given internal limitations and external demands. 

\begin{figure*}[t] 
  \centering
  \includegraphics[width=1\textwidth]{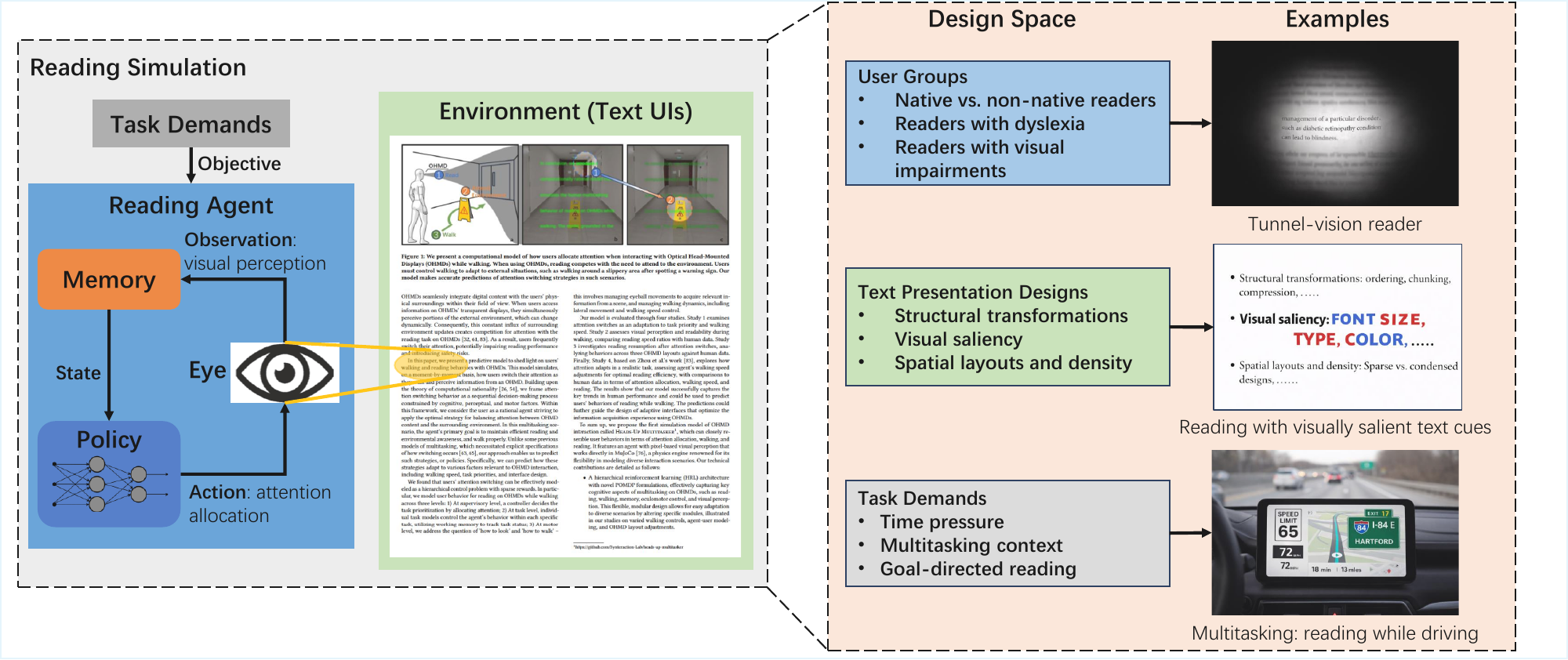}
  \captionsetup{width=1\textwidth}
  \caption{
  \textbf{Resource-rational reading models enable principled optimization of text user interfaces across a wide design space.}
  (\textit{Left}) A reading agent interacts with text-based user interfaces by allocating visual attention over time, guided by perceptual input, memory state, task objectives, and an adaptive policy. (\textit{Right}) A design space for augmented reading defined by user groups, text presentation designs, and task demands. The same underlying model can be instantiated to evaluate diverse reading scenarios, illustrated here with three examples: (\textit{top}) a reader with restricted visual field (tunnel vision), (\textit{middle}) reading supported by visually salient text cues (e.g., font size, type, and color), and (\textit{bottom}) multitasking contexts such as reading while driving from an in-vehicle display. Together, the model enables principled comparison and optimization of text UI designs across users, tasks, and environments.
  }
  \Description{}
  \label{fig:model and design space} 
\end{figure*}

As illustrated in Figure~\ref{fig:model and design space}, a reader is modeled as an agent that interacts with a text environment through perception and action. The agent selects reading actions, such as where to fixate, whether to continue or regress, and when to stop based on partial and uncertain observations. Cognitive constraints, including perceptual noise, memory limitations, and attentional capacity, are explicitly represented, inducing uncertainty and trade-offs that shape reading behavior. Reading strategies are derived through optimization in simulation, rather than by directly fitting human data, grounding the resulting behavior in explicit cognitive assumptions.


This simulation-based, resource-rational approach offers several advantages for HCI research~\cite{murray2022simulation,oulasvirta2022computational}. It provides a generative and explainable account of reading behavior, where eye movements, reading speed, regressions, and comprehension emerge from optimization under constraints rather than from heuristics or post hoc data fitting~\cite{bai2025resource}. Because policies are learned in simulation, the approach is also less data-intensive than purely data-driven models while remaining quantitatively testable~\cite{sutton1998reinforcement}. Prior work shows that such models reproduce key empirical phenomena in reading and multitasking, including attention switching, resumption costs, and performance degradation under time pressure or divided attention~\cite{bai2024heads,bai2025resource,lingler2024supporting}. Crucially for augmented reading, resource-rational models support systematic adaptation across contexts and individuals~\cite{bai2025resource}. Task demands, environmental constraints, and display properties can be modeled through changes in rewards or constraints~\cite{bai2025resource,bai2024heads,lingler2024supporting,jokinen2020adaptive}, while individual differences can be captured by adjusting internal parameters such as memory capacity or lexical efficiency~\cite{chandramouli2024workflow,kangasraasio2019parameter}. This makes simulated readers suitable not only for explaining behavior, but also for evaluating and optimizing alternative text presentations, enabling principled exploration of adaptive designs beyond static layouts or heuristic rules~\cite{bai2025resource}.


\begin{figure*}[t] 
  \centering
  \includegraphics[width=1\textwidth]{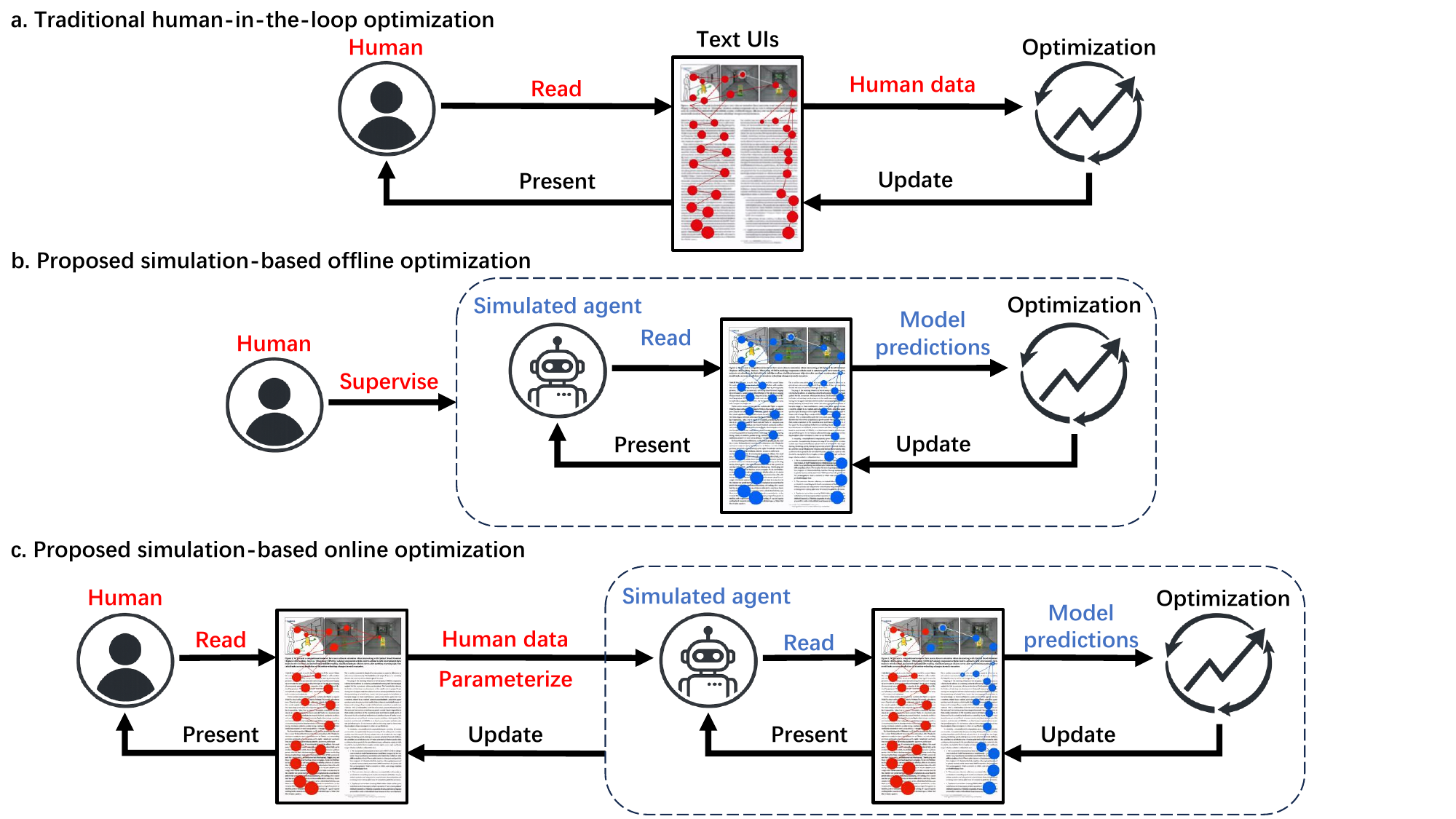}
  \captionsetup{width=1\textwidth}
  \caption{
  \textbf{Optimization paradigms for augmented reading.}
  (a) Human-in-the-loop optimization. Traditional text UI design relies on human readers to generate behavioral data and feedback, making optimization costly and difficult to scale.
  (b) Simulation-based offline optimization. A simulated reading agent represents human readers to evaluate and optimize candidate text UIs, enabling systematic exploration of the design space without repeated user studies.
  (c) Simulation-based online optimization. Human interaction data are used to initialize a personalized simulated agent that continues reading on the user’s behalf, supporting real-time evaluation and adaptive UI optimization during ongoing reading.
  }
  \Description{}
  \label{fig:optimization paradigms} 
\end{figure*}

\section{Augmented Reading as an Optimization Problem}

\subsection{Definitions}
\paragraph{\textbf{Augmented reading.}}
We define augmented reading as text user interfaces that adapt presentation to support readers under varying tasks, contexts, and capabilities. Rather than treating text as static, augmented reading systems dynamically modify layout, content, or visual properties to improve task performance. Importantly, ``good reading'' is not defined by a single metric such as speed or readability, but by how effectively readers achieve their goals under cognitive and environmental constraints. Reading involves fundamental trade-offs between comprehension, effort, time, and attention~\cite{rayner1998eye,bai2025resource}. Augmented reading therefore requires adapting text presentation based on the reader’s situation and task demands, with the goal of supporting task-relevant comprehension as efficiently as possible.

\paragraph{\textbf{Reading as a resource-rational optimization problem.}}
Resource rationality frame reading as adaptive decision-making under cognitive and environmental constraints. Readers allocate limited resources—such as visual attention, working memory, and time—to maximize task utility, for example comprehension or information extraction~\cite{bai2024heads,bai2025resource}. These decisions unfold hierarchically, from eye-movement control to higher-level choices about pacing, rereading, or terminating reading. From this perspective, augmented reading can be formulated as an optimization problem: given a reader, a task, and a context, how should text be presented to maximize expected utility under resource constraints? This framing shifts attention from individual interface features to the objectives they serve, providing a principled basis for evaluating and optimizing alternative text presentations.

\subsection{Design Space in Augmented Reading}

Augmented reading systems operate over a rich design space that shapes how text is perceived, processed, and acted upon. As illustrated in Figure~\ref{fig:model and design space}, this space spans (i) \emph{user groups} (e.g., readers with different perceptual or cognitive capabilities), (ii) \emph{text presentation designs} (e.g., ordering, chunking, compression, visual saliency, and layout density), and (iii) \emph{task demands} (e.g., time pressure, multitasking, or goal-directed reading).

Resource-rational reading models provide a principled way to navigate this space. By explicitly modeling how design factors interact with human cognitive constraints, the same underlying reading model (Figure~\ref{fig:model and design space}, left) can be instantiated to evaluate diverse scenarios (Figure~\ref{fig:model and design space}, right), enabling systematic comparison and optimization of text UIs that would be difficult to achieve through manual design alone.

\subsection{Resource-Rational Reading Models as Evaluators}

Because resource-rational reading models explicitly characterize how human readers adapt to cognitive and environmental constraints, they can be repurposed as evaluators for augmented reading design. Rather than predicting interface adaptations directly from data, as in many data-driven approaches~\cite{bolliger2023scandl,Bolliger2025ScanDL2,deng2023eyettention}, a simulated reader is exposed to candidate text presentations and used to estimate their expected utility in terms of comprehension, effort, and time costs. This evaluator-based perspective enables systematic and consistent comparison of alternative designs across a large design space, supporting exploration and optimization beyond what is feasible through manual iteration or user testing alone~\cite{murray2022simulation}. By adjusting internal model parameters such as memory capacity, perceptual noise, or lexical efficiency, the same simulated reader can approximate different user profiles, including readers under extreme multitasking or atypical cognitive constraints. As a result, resource-rational models support reasoning about individual differences and edge cases, enabling more inclusive, robust, and cognitively grounded augmented reading systems.

\section{Illustrative Scenarios and Design Implications}

\begin{figure*}[t] 
  \centering
  \includegraphics[width=1\textwidth]{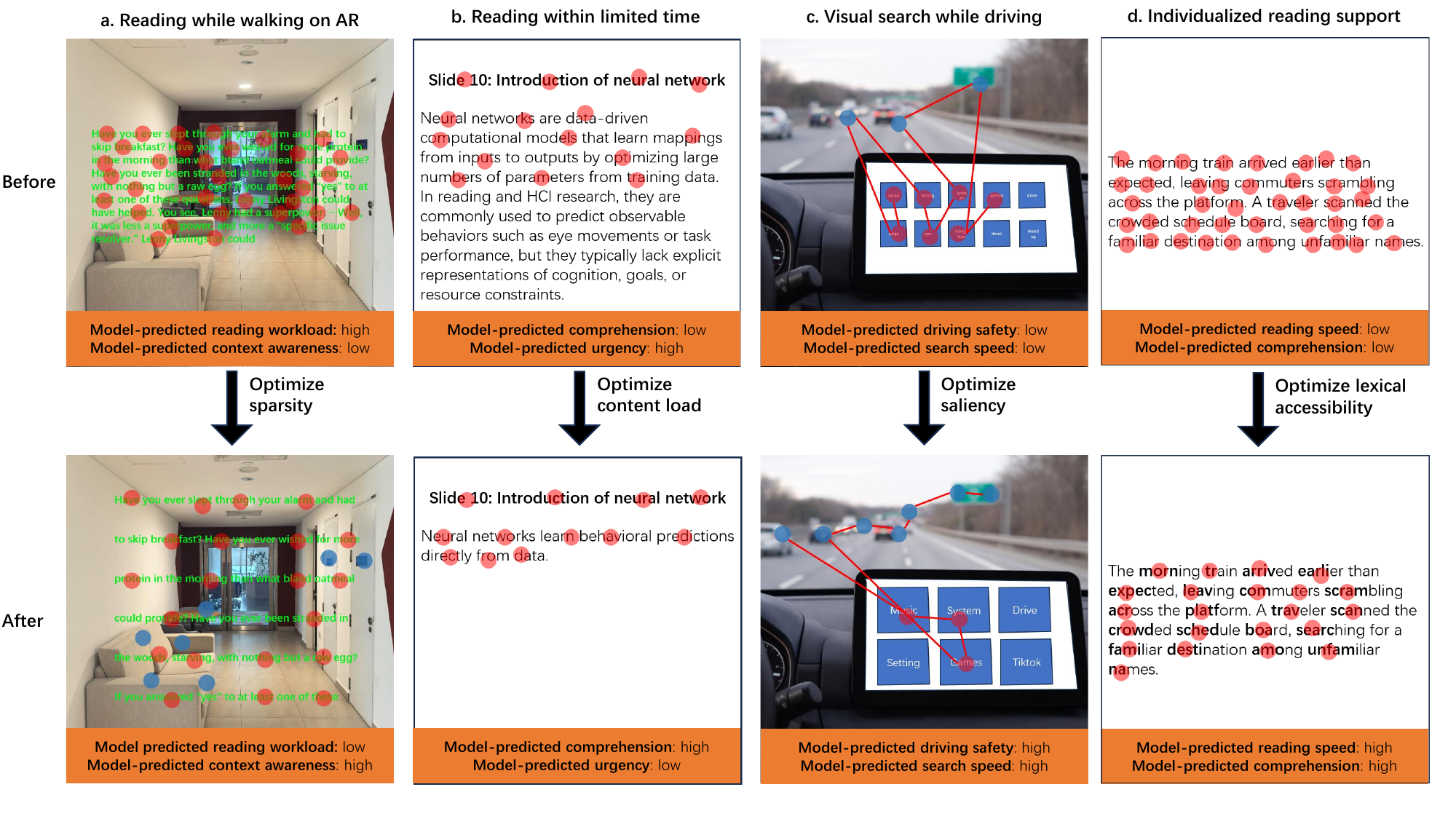}
  \captionsetup{width=1\textwidth}
  \caption{
  \textbf{Gallery of simulation-based optimization for augmented reading.}
  Three scenarios illustrate how a simulated reading agent predicts human-like reading behavior (red dots indicate fixations on text; blue dots indicate attention to the surrounding environment) and task performance (summarized by model-predicted metrics in orange boxes) under different contexts, and how these predictions guide UI optimization. Across (a) reading while walking on AR, (b) time-limited reading, and (c) visual search while driving, the model identifies bottlenecks in workload, comprehension, or safety. Targeted design interventions, such as increasing layout sparsity, reducing content load, or enhancing visual saliency lead to improved predicted reading behavior and task performance after optimization. (d) Individualized reading support demonstrates how reader-specific simulation reveals inefficient lexical access (e.g., slow reading speed and dense fixations), enabling personalized adaptations such as enhanced lexical saliency to improve predicted reading speed and comprehension.
  }
  \Description{}
  \label{fig:simulation cases} 
\end{figure*}

We outline two complementary optimization paradigms for augmented reading based on resource-rational reading models: offline~\cite{murray2022simulation,junior2024agentforge} optimization and online~\cite{lindlbauer2019context} optimization (Figure~\ref{fig:optimization paradigms}). The two paradigms differ in whether humans are inside or outside the optimization loop, but share the same principle: using simulation to guide principled, scalable interface adaptation.

\subsection{Offline Optimization}

In offline optimization, simulated readers are used to evaluate and refine text interfaces prior to deployment (Figure~\ref{fig:optimization paradigms}~b). Given a candidate design, such as a layout, summarization level, or saliency scheme, the simulated agent reads under specified task demands (e.g., time pressure or comprehension goals) and produces predicted behavioral and performance metrics, including fixation patterns, reading time, and comprehension. These predictions reveal cognitively demanding regions, such as passages associated with dense fixations or frequent regressions, enabling designers to target revisions (e.g., restructuring content, reducing load, or increasing saliency). Revised designs can be iteratively re-evaluated in simulation, supporting systematic exploration of large design spaces without repeated user studies. By adjusting model parameters, the same pipeline can approximate different reader populations or capabilities.




\subsection{Online Optimization}

In online optimization, a simulated reader supports real-time adaptation during ongoing reading (Figure~\ref{fig:optimization paradigms}~c). Partial observations of a user’s interaction, such as eye movements, reading speed, or task progress are used to initialize or update the internal state of the simulated agent. Conditioned on this interaction history, the agent predicts near-future reading behavior and resource demands under the current text UI. These predictions enable in-situ interface optimization while reading is still in progress. The simulated agent iteratively evaluates candidate adaptations on the user’s behalf, selecting text presentations that better balance comprehension, effort, and time under the current context. By grounding adaptation in an explicit model of human cognition, this paradigm supports dynamic, personalized UI updates that account for high-level cognitive factors rather than relying on surface heuristics or reactive rules.

\subsection{Design Implications}


Together, these optimization paradigms enable simulation-based design workflows for augmented reading. As illustrated in Figure~\ref{fig:simulation cases}, model predictions can identify task-specific bottlenecks (e.g., workload, comprehension, or safety) and guide targeted design interventions that improve predicted reading and task performance across contexts. This positions resource-rational reading models as practical tools for exploring, comparing, and optimizing adaptive text interfaces.


\section{Conclusion}
We framed augmented reading as a simulation-based optimization problem grounded in resource-rational models of human reading. By using cognitively grounded simulated readers as evaluators, both offline and online optimization pipelines can systematically improve text interfaces with respect to comprehension, effort, and time under diverse contexts and user profiles. This perspective enables adaptive, explainable, and scalable augmented reading design that goes beyond heuristics, large-scale data collection, or repeated human testing.

\begin{acks}
A.~O. was supported by the European Research Council (ERC; Grant No.~101141916) and the Research Council of Finland (Grant Nos.~328400, 345604, 341763, and 357578). 
S.~Z. was supported by the City University of Hong Kong (Grant No.~9610677). 
\end{acks}

\bibliographystyle{ACM-Reference-Format}
\bibliography{reference}

\end{document}